# Single-shot d-scan technique for ultrashort laser pulse characterization using transverse second-harmonic generation in random nonlinear crystals

Francisco J. Salgado-Remacha,[1,†] Benjamín Alonso,[2,†] Helder Crespo,[3,4] Crina Cojocaru,[5] Jose Trull,[5] Rosa Romero,[4] Miguel López-Ripa,[2] Paulo T. Guerreiro,[4] Francisco Silva,[4] Miguel Miranda,[3,4,6] Anne L'Huillier,[6] Cord L. Arnold,[6] and Íñigo J. Sola[2,*]

[1]*Departamento de Física Aplicada, Facultad de Ciencias, Universidad de Zaragoza, C/ Pedro Cerbuna 12, 50009 Zaragoza, Spain*
[2]*Grupo de Aplicaciones del Láser y Fotónica (ALF), Departamento de Física Aplicada, University of Salamanca, Plaza de la Merced s/n 37008 Salamanca, Spain*
[3]*IFIMUP and Departamento de Física e Astronomia, Universidade do Porto, Rua do Campo Alegre 687, 4169-007 Porto, Portugal*
[4]*Sphere Ultrafast Photonics, Rua do Campo Alegre, n.º 1021, Edifício FC6, 4169-007 Porto, Portugal*
[5]*Departament de Física, Universitat Politècnica Catalunya, Terrassa, 08222, Spain*
[6]*Department of Physics, Lund University, P.O. Box 118, SE-221 00 Lund, Sweden*
*\*Corresponding author: ijsola@usal.es*

*We demonstrate a novel dispersion-scan (d-scan) scheme for single-shot temporal characterization of ultrashort laser pulses. The novelty of this method relies on the use of a highly dispersive crystal featuring antiparallel nonlinear domains with a random distribution and size. This crystal, capable of generating a transverse second-harmonic signal, acts simultaneously as the dispersive element and the nonlinear medium of the d-scan device. The resulting in-line architecture makes the technique very simple and robust, allowing the acquisition of single-shot d-scan traces in real time. In addition, the technique can be further simplified by avoiding the need of dispersion pre-compensation. The retrieved pulses are in very good agreement with independent FROG measurements. We also apply the new single-shot d-scan to a TW-class laser equipped with a programmable pulse shaper, obtaining an excellent agreement between the applied and the d-scan retrieved dispersions.*

High-power ultrashort laser pulses are increasingly important tools in different fields of research and technology (e.g., femtochemistry, biology, atomic and molecular physics, and particle acceleration, among others). Most of these applications need very precise characterization of the laser pulse parameters, namely the pulse duration, spectrum and spectral phase. For more than 20 years, the most common methods for complete laser pulse characterization (amplitude and phase) have been Frequency Resolved Optical Gating (FROG) [1] and Spectral Phase Interferometry for Direct Electric-field Reconstruction (SPIDER) [2]. More recently, the dispersion-scan (d-scan) technique [3] was proposed for the complete measurement of ultrashort laser pulses. In this method, variable dispersion is applied to the pulse prior to generating a given nonlinear signal. The resulting two-dimensional dispersion scan trace encodes the spectral phase of the original pulse. Pulse retrieval is performed via a holistic algorithm that iteratively adjusts a parametrized pulse spectral phase in order to match the calculated and the measured d-scan traces [3,4].

One common experimental d-scan approach uses a set of chirped mirrors to add negative dispersion to the pulses and a pair of glass wedges to perform the dispersion scan, by progressively inserting one or both wedges while crossing the optimum compression point. The most usual nonlinear process is second-harmonic generation (SHG) in thin crystals [3,4], in order to maximize the phase matching bandwidth. The advantage of the wedge-pair-based scanning relies on its simplicity, low cost and the fact that it is particularly well-suited for extremely short pulses in the few- to single-cycle regime [5,6]. On the other hand, the low amount of dispersion introduced by wedges limits its use to very short duration pulses compared to other methods [1,2].

The d-scan technique is very compact and robust, since its single-beam in-line setup obviates the need of pulse splitting and temporal delays. However, the common implementation of the technique, which requires changing the wedge insertion, cannot easily be adapted to single-shot measurements. This can be a major drawback when characterizing laser systems where the pulse parameters may vary from one pulse to the next. Indeed, high-power lasers (multi-TW-class and PW-class systems) are increasingly common [7] and due to their low repetition rate and pulse-to-pulse variations, single-shot methods are crucial for their characterization. In order to adapt the d-scan technique to this scenario, single-shot architectures have been proposed [8,9]. These approaches are wedge-based, with a limited amount of introduced dispersion, which means that pulses with Fourier-limited durations larger than 30-40 fs are difficult to measure.

In this work, we present a new approach and setup for a d-scan technique capable of performing single-shot measurements, based on random nonlinear crystals exhibiting broadband *transverse second harmonic generation* (TSHG) [10,11]. First, we present the single-shot d-scan experimental setup, highlighting the properties of the nonlinear medium. Then we demonstrate the system performance, including further setup simplification by avoiding the use of the chirp pre-compensation, and comparing the single-shot d-scan measurements with those obtained by another technique (FROG). Finally, the method is applied to the characterisation of pulses from a TW-class laser.

The novelty of the proposed method relies on the implementation of a highly dispersive random nonlinear crystal, replacing the phase-matched thin nonlinear crystal used in a typical autocorrelator or d-scan setup. Some of the as-grown ferroelectric crystals (i.e., with no electric field applied during its growth), such as Strontium Barium Niobate (SBN) [12], naturally exhibit a random-sized distribution of needle-like antiparallel ferroelectric domains, with typical dimensions below 1 μm, oriented parallel to the optical axis. While the reversed orientation of domains corresponds to an inversion of the quadratic susceptibility sign, the refractive index of the crystal remains practically homogeneous [13]. Such nonlinear domain distribution creates a continuous set of reciprocal lattice vectors with different magnitudes and orientations within the plane perpendicular to the optical axis, providing phase mismatch compensation for SHG over very broad wavelength and angular ranges [10,11]. Consequently, the SHG signal is generated in the whole plane perpendicular to the optical axis of the crystal, thus being a transverse second-harmonic generation (TSHG) process. These unique emission properties make possible the detection of the SHG signal in the direction transverse to pulse propagation. This configuration has been recently used in transverse auto- [14] and cross- [15] correlation schemes where the nonlinear signal was recorded directly from the top surface of the crystal, following the evolution of the TSHG signal along the length of the crystal, which is not possible in any other conventional auto- or cross-correlation scheme. While transverse autocorrelation cannot provide direct information about the spectral phase, TSHG is well-suited to d-scan since the medium performs simultaneously the two functions needed for the technique. First, since phase matching is no longer an issue, the TSHG crystals can be long (some mm or cm along the beam propagation direction) [14]. Therefore, the material dispersion along the beam propagation can be used as an effective dispersion scanning element. Furthermore, a TSHG signal is produced at each propagation plane within the material (i.e., at each spectral phase addition). Thus, the unique characteristics of TSHG crystals enable designing a new single-shot d-scan scheme, where the nonlinear medium acts as both the dispersion scanning module and the nonlinear signal generator. Fig. 1 shows the experimental setup. The pulse to be analyzed passes first through a pre-compensation negative dispersion setup, in our case dispersive chirped mirrors, DCMs (32 bounces off Layertec chirped mirrors, providing a nominal group delay dispersion (GDD) of -40 ± 15 $fs^2$ per bounce over the 700-900 nm spectral range). This pre-compensation setup provides spectral phase compensation and pulse compression within the TSHG crystal and is required to generate a complete d-scan trace. A half-wave plate is used to rotate the linear polarization of the beam to the direction optimizing the nonlinear process. Then, the pulse enters the nonlinear medium, in our case a 1-cm-long SBN crystal with the optical axis oriented in the plane normal to the beam propagation. As the pulse propagates, it acquires the spectral phase due to the material's dispersion (positive chirp in the near-IR spectral range) and produces TSHG at the same time. A lens images the TSHG emission onto the input slit of an imaging spectrometer (a Czerny-Turner type spectrometer, Shamrock 500i from Andor), resulting in a spectrally resolved image of the TSHG, i.e., a single-shot d-scan trace. As a light source, we used an optical parametric amplifier (OPA) (TOPAS from Light Conversion), pumped by a chirped pulse amplification (CPA) laser (Spitfire ACE from Spectra Physics) delivering pulses with a full-width-at-half-maximum (FWHM) duration of around 64 fs and 1.6 mJ of energy per pulse.

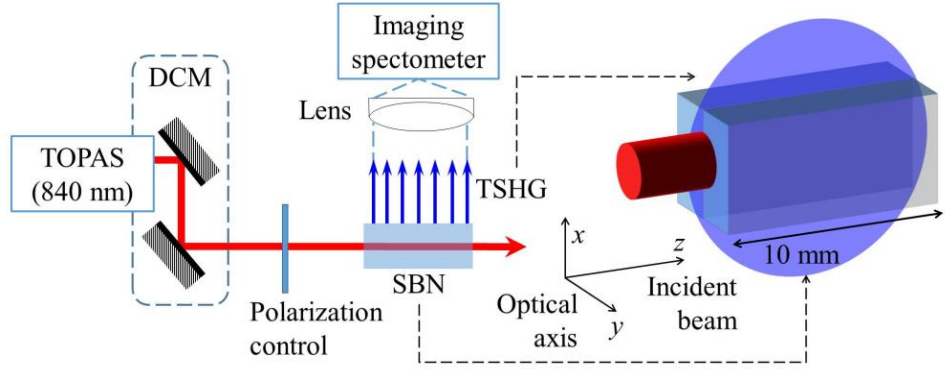

*Fig. 1. Schematic of the single-shot d-scan pulse characterisation setup (see text for more details).*

A key parameter in the measurement is the dispersion of the TSHG crystal, since it defines one of the axes of the d-scan trace. This requires independent characterization of the optical properties of the crystal. An alternative is to use the so-called self-calibrating d-scan reconstruction algorithm [16], which allows retrieving both the spectral phase of the pulses and the unknown nominal dispersion of the dispersion scan system (e.g., wedge material and insertion, prism or grating compressor parameters, etc.). In the present case, the varying dispersion is a result of the different depths of propagation of the pulses inside the bulk SBN crystal. The dispersion of the SBN crystal was modelled by considering the contributions of the group velocity dispersion (GVD) and the third-order dispersion (TOD).

In order to validate the setup and the technique, we performed single-shot d-scan measurements of ultrashort laser pulses and compared the results with measurements performed with a home-built FROG set-up based on non-collinear SHG in a thin (20-μm-thick) BBO crystal. Figure 2 shows the results of single-shot d-scan measurements, including the measured d-scan trace (Fig. 2a), the retrieved trace (Fig. 2b), the spectral intensity and retrieved phase (Fig. 2c), and the temporal intensity and phase (Fig. 2d) of the pulses prior to the DCMs. This implies retrieving the pulses at the input plane of the crystal and extracting the spectral phase provided by the 32 bounces off the DCMs, as described below. The retrieved pulses have a duration of 46 fs (FWHM). The SBN dispersive properties obtained from the self-calibrating algorithm are GVD=430 $fs^2$/mm, in very good agreement with the expected value of GVD=439.5 $fs^2$/mm at 840 nm [17], and the value of the TOD is quite small, with no noticeable effect on the pulse propagation, as observed in previous work [14]. In order to validate these results, we changed the input pulse spectral phase by introducing known amounts of dispersion and recovered its values from the d-scan measurements. By varying the number of bounces off the DCMs, apart from the above-mentioned case of 32 bounces (e.g., by using 0, 6, 12, 24 and 48 reflections), the spectral phase introduced by the DCMs can be extracted. The measured GDD per bounce is -37 ± 4 $fs^2$, in very good agreement with the DCM nominal value previously mentioned.

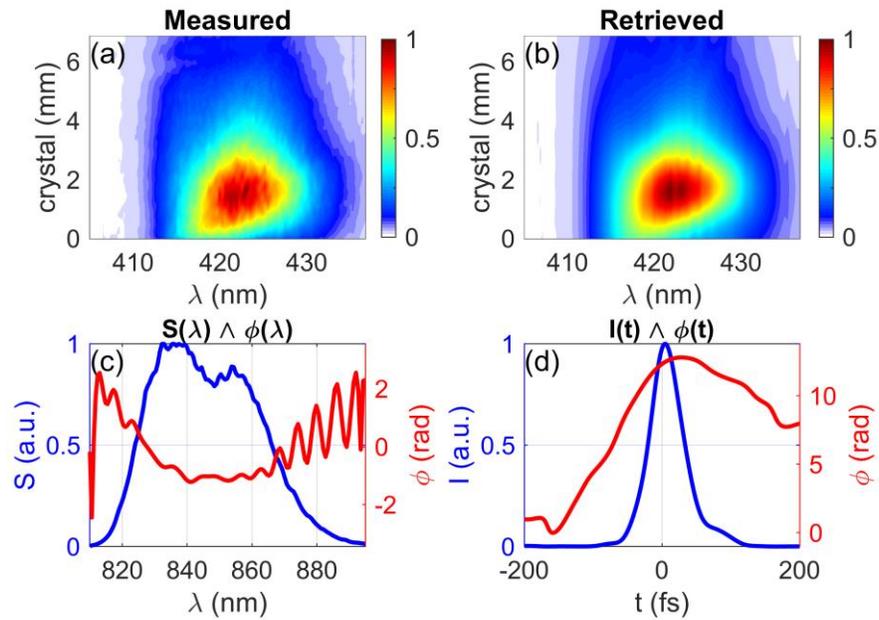

*Fig. 2. Single-shot d-scan measurement of pulses from an OPA tuned to 840 nm after 32 bounces in DCMs: experimental d-scan trace (a), retrieved trace (b), spectral intensity (blue) and retrieved phase (red) (c) and retrieved temporal intensity (blue) and phase (red) (d). The spectral and temporal reconstructions shown at (c) and (d) respectively correspond to the pulse at the input of the device (i.e., prior to the DCMs), obtained by subtracting the dispersion introduced by the 32 bounces off the DCMs to the retrieved pulse at the input of the TSHG crystal.*

Among the cases using different numbers of bounces in the DCMs, the one using no DCMs at all (Fig. 3) is especially interesting. The experimental d-scan trace (Fig. 3a) features contributions only due to addition of positive chirp to the input pulse, but no negative chirp, since there is no chirp pre-compensation. The convergence of the self-calibrating d-scan retrieval of the trace (Fig. 3b) is good, in agreement with previous theoretical work where retrieval from partial d-scan traces allowed reconstructing the pulses with low error [18]. Figures 3c and 3d show the spectral intensity (blue line) and retrieved phase (red line), and the temporal intensity (blue line) and phase (red line), respectively. The retrieved pulse has a duration of 44 fs (FWHM). In Fig. 4d the retrieved pulse for this 0-bounces case, shown in Fig. 3d, and the 32-bounces case, shown in Fig. 2d, are compared, presenting a good agreement not only in duration but also in temporal shape. Therefore, it is possible to simplify further the single-shot d-scan setup by skipping the negative pre-compensation DCMs. This allows extending the spectral range of operation, since it obviates the need of DCMs, which are designed for a given input pulse bandwidth.

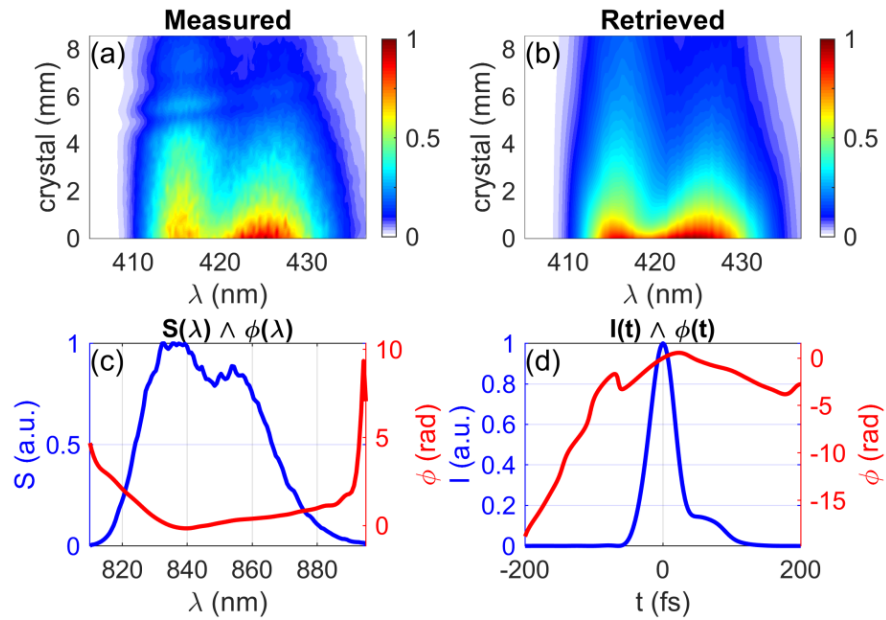

*Fig. 3. Single-shot d-scan measurement of pulses from an OPA tuned at 850 nm without dispersion pre-compensation (0 bounces in DCMs): d-scan experimental trace (a), retrieved trace (b), spectral intensity (blue) and retrieved phase (red) (c) and retrieved temporal intensity (blue) and phase (red) (d).*

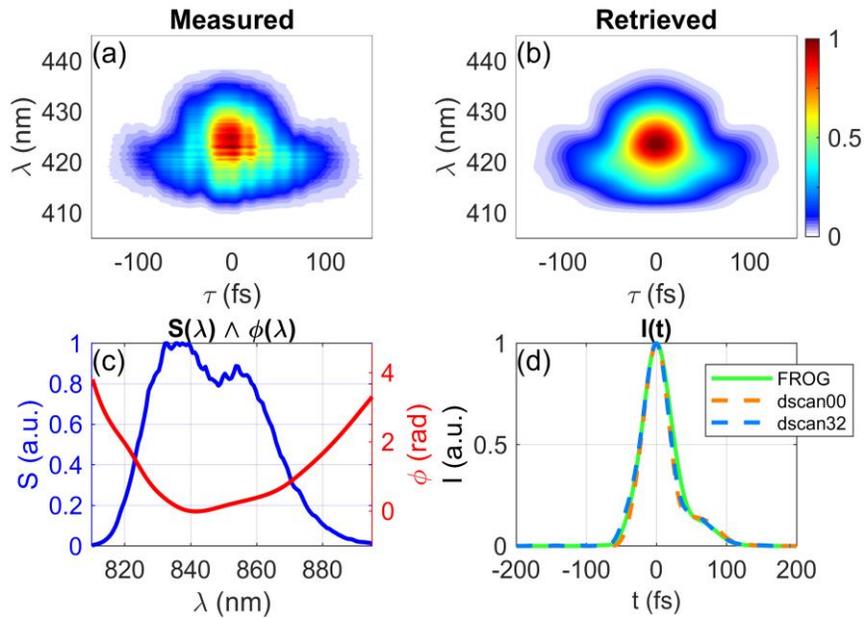

*Fig. 4. Comparison between single-shot d-scan and FROG: Experimental (a) and retrieved FROG traces (b), retrieved (blue) and measured (red) spectral intensity (c) and temporal intensity retrieved from FROG (d, green), compared with the corresponding pulse retrieved from the d-scan measurement using 32 DCM bounces (d, dashed blue) and no DCM bounces (d, dashed orange).*

FROG measurements of the input pulse before the DCMs are presented in Fig. 4, giving a pulse duration of 47 fs (FWHM). Fig. 4d compares the FROG measurement and the d-scan measurements before the 32 DCM bounces (Fig. 2d), and for the 0 DCM bounces case (Fig. 3d). The agreement between the three retrievals is very good in both pulse duration (47 fs from FROG vs. 46 fs from 32 DCM-bounce d-scan vs. 44 fs from 0 DCM-bounce d-scan) and shape, validating both single-shot d-scan retrievals (i.e., performed with 32 DCM bounces and without any bounces).

The single-shot d-scan system was also used to measure laser pulses produced by a 40 TW CPA laser system at a repetition frequency of 10 Hz with 35 fs pulse duration. This system includes an acousto-optic programmable dispersive filter (DAZZLER) that enables adjusting the phase, which we used to add a number of known GDD values. The experimental GDD variation (dots) is obtained from the retrieved spectral phase and presents very good agreement with the nominal change of the GDD (dashed line), as summarized in Fig. 5.

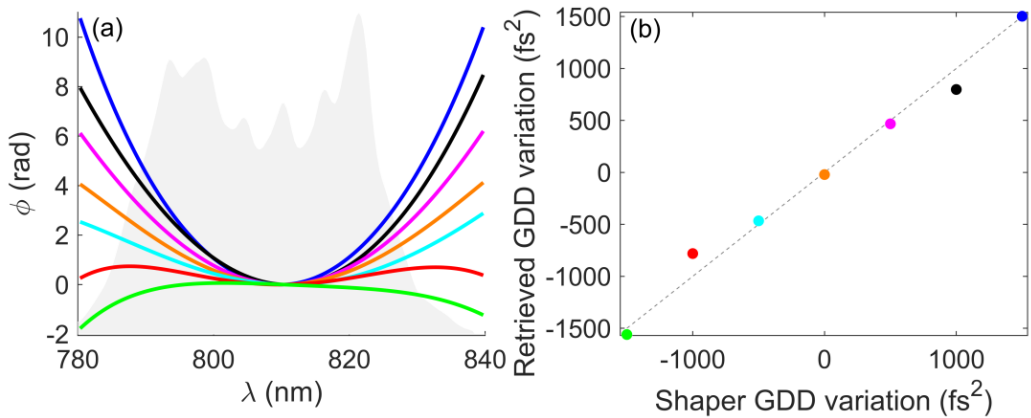

*Figure 5. D-scan phase reconstruction for different experimental GDD values applied by a pulse shaper: (a) experimental spectral intensity (shadow) and retrieved spectral phase (coloured lines) and (b) retrieved relative GDD values from the single-shot d-scan phase (coloured dots) and nominal GDD value applied by the pulse shaper (dashed line). The dots in (b) correspond to the same colour spectral phase in (a). The GDD values are relative to the pulse marked in orange.*

In conclusion, we presented and demonstrated a single-shot d-scan system, based on a new architecture and nonlinear random crystals, able to measure ultrashort laser pulses using a very simple and robust configuration. Our system is easy to align and especially well-suited to the characterisation of high-power, low repetition rate ultrashort lasers and other systems where pulses can change rapidly from shot-to-shot. We describe two possible configurations (with and without negative dispersion pre-compensation) and validate our results by comparing with another well-established pulse characterisation technique. The application of the self-calibrating d-scan algorithm conveniently provides automatic calibration of the dispersion characteristics of the used SBN crystal. The single-shot d-scan system was also used for the measurement of a high-power 40 TW CPA laser system with good agreement between the retrieved GDD values and the nominal ones added to the pulses.

Because of the SBN crystal dimensions and very favourable optical characteristics (large transparency and phase matching spectral window as well as high dispersion) [14], this experimental setup is able to measure pulses with spectra corresponding to Fourier-limited durations in the range of 15 fs to 60 fs in the 800-900 nm spectral region. It may be possible to characterise shorter pulses with broader spectra centred at longer wavelengths within the SBN transparency window. In addition, the application range may be extended towards longer Fourier-limited pulse durations, by simply increasing the medium length and hence the amount of dispersion that can be applied to the pulse.

We should point out that, in contrast to other single-shot techniques [8,9,19], the beam does not have to be spatially homogeneous, and the diagnostic can be used to measure the pulse at different sampling positions within the beam.

**Funding.** Junta de Castilla y León (SA046U16 and SA287P18), Spanish MINECO (FIS2017-87970-R, EQC2018-004117-P), US Army Research, Development and Engineering Command (RDECOM) (W911NF-16-1-0563 ), Fundação para a Ciência e a Tecnologia (FCT) ('UltraGraf' M-ERA-NET2/0002/2016, 'UltraGraf' M-ERA-NET4/0004/2016PTDC/FIS-OTI/32213/2017, UIDB/04968/2020), Compete 2020, in its FEDER component (Network of Extreme Conditions Laboratories - NECL, NORTE-01-0145-FEDER-022096), PT2020 (program 04/SI/2019 Projetos I&D industrial à escala grant no. 045932), Swedish Research Council, European Research Council (Proof of Concept SISCAN-78992)

[†] These authors contributed equally to this Letter.